\documentclass[%
 reprint,
 amsmath,amssymb,
 aps,
]{revtex4-1}

\usepackage{graphicx}% Include figure files
\usepackage{dcolumn}% Align table columns on decimal point
\usepackage{bm}% bold math
\begin{document}

\preprint{APS/123-QED}

\title{Threshold Effects in the Decay of Heavy $b'$ and $t'$ Quarks}

\author{Yuan Chao$^{a}$, Kai-Feng Chen$^{a}$, Shing-Kuo Chen$^{a}$,
 Wei-Shu Hou$^{a,b}$, Bo-Yan Huang$^{a}$, and Yeong-Jyi Lei$^{a}$}
 \affiliation{$^{a}$Department of Physics, National Taiwan University,Taipei, Taiwan 10617\\
$^{b}$National Center for Theoretical Sciences, National Taiwan University, Taipei, Taiwan 10617}%Lines break automatically or can be forced with \\

%\date{\today}% It is always \today, today,
             %  but any date may be explicitly specified

\begin{abstract}
A sequential fourth generation is still viable, but the $t'$ and
$b'$ quarks are constrained to be not too far apart in mass.
The $t' \to bW$ and $b' \to tW$ decay channels are still being pursued
at the Tevatron, which would soon be surpassed by the LHC.
We use a convolution method with up to five-body final state to study
$t'$ and $b'$ decays. We show how the two decay branches for $m_{b'}$
below the ${tW}$ threshold, $b' \to tW^*$ and $t^*W$, merge with $b' \to tW$
above the threshold. We then consider the heavy-to-heavy transitions
$b' \to t^{\prime(*)}W^{(*)}$ (or $t' \to b^{\prime(*)}W^{(*)}$),
as they are not suppressed by quark mixing. We find that,
because of the threshold sensitivity of the branching fraction of
$t' \to b'W^*$ (or $b' \to t'W^*$), it is possible to measure the
strength of the CKM mixing element $V_{t'b}$ (or $V_{tb'}$),
especially when it is rather small.
We urge the experiments to pursue and separate the $t' \to b'W^*$ (or $b' \to t'W^*$)
decay in their search program.
\begin{description}
%\item[Usage]
% Secondary publications and information retrieval purposes.
\item[PACS numbers]
14.65.Jk, 13.30.Eg, 13.38.Be
%\item[Structure]
% You may use the \texttt{description} environment to structure your abstract; use the optional argument of the \verb+\item+ command to give the category of each item.
\end{description}
\end{abstract}

\pacs{Valid PACS appear here}% PACS, the Physics and Astronomy
                             % Classification Scheme.
%\keywords{Suggested keywords}%Use showkeys class option if keyword
                              %display desired
\maketitle

%\tableofcontents

\section{\label{sec:Intro}INTRODUCTION\protect\\}

As first pointed out by Kobayashi and
Maskawa~\cite{Kobayashi:1973fv} (KM), if Nature possesses three
generations of quarks, then there would exist an irremovable $CP$
violating phase in the charge current. With the emergence of the
$\tau$ lepton and the $b$ quark, this picture quickly became the
basis for the flavor part of the Standard Model (SM). Remarkably,
this KM theory can explain all phenomena observed so far,
culminating in the confirmation of the $CP$ phase by the B factory
experiments in 2001~\cite{PDG}. But the 3 generation SM cannot be a
complete theory even in regards $CP$ violation, as it falls far
short from what is needed for the matter dominance of our
Universe. However, extending to a fourth generation of quarks,
which does not add any new dynamics to the SM, one could attain
enough $CP$ violation for matter dominance~\cite{Hou:2008xd}. In
any case, despite the usual prejudice, a fourth generation of
quarks is still quite viable~\cite{Kribs:2007nz,Holdom:2009rf}.
Through out this report, we use $t^\prime$ and $b^\prime$ to
represent the sequential up- and down-type fourth generation quarks,
respectively.

The Tevatron has held the energy frontier for two decades,
which was surpassed by the LHC at the end of 2009.
Utilizing the collision of $p$ and $\bar p$ beams at
$\sqrt{s} = 1.96$ TeV, the CDF and D$0$ experiments have performed
direct searches~\cite{PDG} for the fourth generation quarks.
The best limits depend on the search channel.
For the search of a top-like heavy quark, i.e. $t' \to qW$,
CDF analyzed 4.6~fb$^{-1}$ collected data, giving a mass limit~\cite{CDFt}
of $m_{t^\prime} >$ 335 GeV at 95\% confidence level (CL).
This has been updated recently to 5.6~fb$^{-1}$, with or without
tagging for a $b$-quark jet. The limit obtained~\cite{Ivanov}
for $t' \to bW$ is at 358 GeV, while for $t' \to qW$ the limit is at 340 GeV,
not much different from the previous result.
The D$0$ experiment has reported recently a similar study
of $t' \to qW$ with 5.3~fb$^{-1}$ collected data, giving a mass limit
of $m_{t^\prime} >$ 285 GeV at 95\% CL~\cite{D0}.
CDF has also searched for pair production of $b'$ quarks, followed by
$b' \to tW$ decay. In the first study based on 2.7~fb$^{-1}$ collected data,
CDF exploited the low background nature of the same-sign dilepton signature
(together with associated jets, one of which $b$-tagged, plus missing transverse energy),
and gave~\cite{Aaltonen:2009nr} the bound of 338 GeV.
A better limit was obtained recently in the lepton plus jets study based on
4.8 fb$^{-1}$ collected data. CDF searched for an excess of events with
an electron or a muon, at least five jets (one tagged as a $b$ or $c$),
and an imbalance of transverse momentum. The observed events
were consistent with background expectations, giving the upper limit of
$m_{b'} > 372$ GeV at 95\% CL~\cite{Aaltonen:2011vr}.

The LHC has seen remarkable performance since 2010.
Already, a search for pair-produced heavy bottom-like quarks in
proton-proton collisions at $\sqrt{s} = 7$ TeV has been reported.
The CMS experiment searched for $b^\prime \overline{b}{}^\prime
 \to t W^- \overline{t}W^+$ with same-sign dileptons, using a
data set of 34~pb$^{-1}$ collected in 2010.
No events were found in the signal region, and
the $b'$ mass range from 255 to 361 GeV was excluded at the 95\%
confidence level~\cite{CMS1}.
For $t'$ (or $b'$) $\to qW$ search, the ATLAS experiment reported recently
a study of dilepton events with 37 pb$^{-1}$ collected in 2010,
using a boosted $W$ approach in a ``colinear mass" variable.
The reported preliminary~\cite{ATLAS1} limit is 270 GeV.
These studies clearly harbinger the passing of the torch from
the Tevatron to the LHC, as far as heavy chiral quark search is concerned.

In this report, we first take closer scrutiny of the $b'\to tW$
decay process to illustrate the width effect involving two unstable
daughters. The decay widths of $b^\prime \to t^{(*)}W^{(*)}$ are obtained
using the convolution method~\cite{BarShalom:2005cf, Altarelli:2000nt,
Calderon:2001qq, Kuksa:2006co, Mahlon:1995co} at tree level.
If the $b^\prime$ mass is below the $tW$ threshold,
then $b^\prime \to tW$ decay is phase space forbidden,
and $b^\prime$ decays via $b^\prime \to tW^*$ and $t^*W$, where either
$t$ or $W$ is off-shell. The former case was missed in a previous
analysis~\cite{Arhrib:2006pm}. Note, however, that each of these
two decay widths would turn into the $b' \to tW$ decay width when
$m_{b^\prime}$ is above the $tW$ threshold. We therefore
investigate how double counting is avoided as the threshold is
approached. In so doing, we elucidate how, for different $b'$ mass
scenarios, the decay rate of $b^\prime \to t^{(*)}W^{(*)}$ can be
effectively 5-body, 4-body, 3-body, or finally, the two-body
$b^\prime \to tW$ process above $tW$ threshold of $255$ GeV.
The method is applied to investigate $b' \to t^{\prime(*)}W^{(*)}$
or $t' \to b^{\prime(*)}W^{(*)}$ decays, depending on mass hierarchy.
In fact, because one expects the $t'$--$b'$ mass splitting to
be less than $M_W$, the dominant process would be $b' \to
t^{\prime}W^{*}$ or $t' \to b^{\prime}W^{*}$ decay, where
the $W$ is virtual. We focus on studying the effect of the
CKM mixing element on $b^\prime$ and $t^\prime$ decays,
especially the near and below threshold behavior.
If the CKM mixing element $|V_{t'b}|$ (or $|V_{tb'}|$) is
small enough, the decay channel $b' \to tW$ (or $t' \to bW$)
would be suppressed. Thus, a measurement of $b' \to t^{\prime}W^*$
(or $t' \to b^{\prime}W^*$) branching fraction would allow one to
in  principle measure the strength of $|V_{t'b}|$ (or $|V_{tb'}|$).
Finally, in an Appendix we compare the calculated decay widths
for $b^\prime \to tW^*$ decay across different thresholds
with results obtained from PYTHIA.

\section{\label{sec:II}Width Effect of Unstable Daughters\protect\\}

\begin{figure}[t]
\centering
%\includegraphics[width=60mm,height=40mm]{plots/twidth.pdf}
%\vspace{30mm}
{\includegraphics[width=60mm]{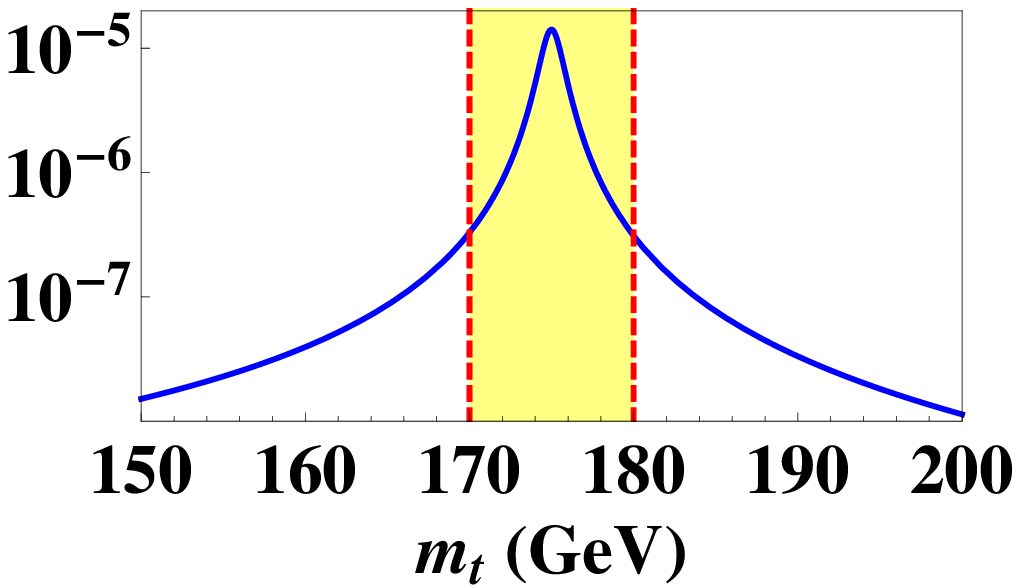}}
\vskip0.35cm
%{\includegraphics[height=47mm]{plots/tbw.eps}}
{\includegraphics[width=60mm]{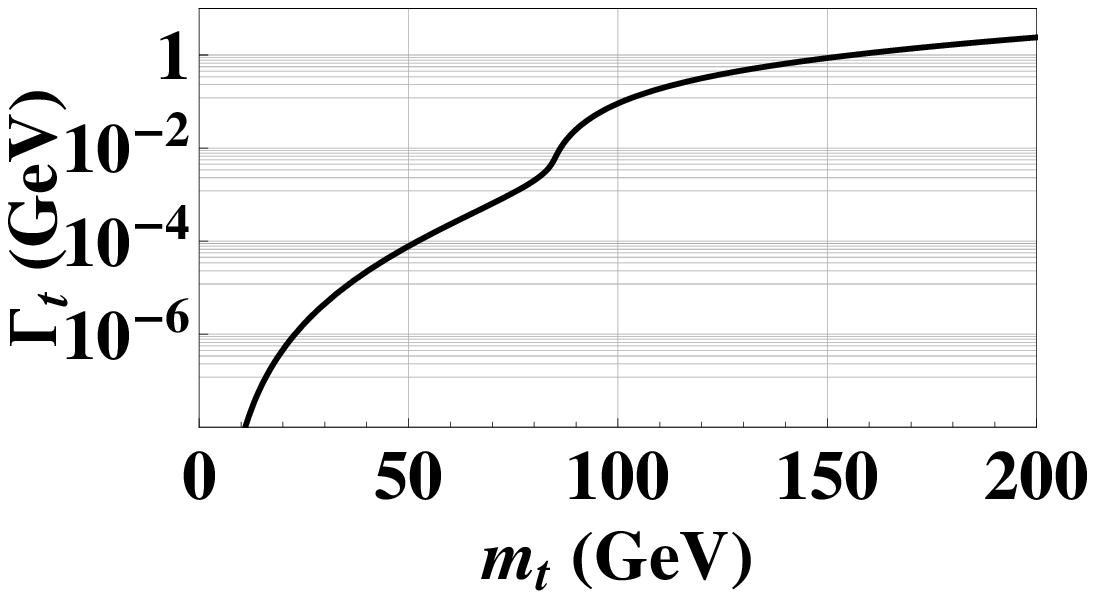}}
\caption
{Upper plot: the top quark resonance width in the form of
a Breit-Wigner distribution; lower plot: the decay width of $t \to
bW$ as a function of $m_{t}$. A clear finite width effect of the
$W$-boson mass threshold is seen.
} \label{tbwwid}
\end{figure}

The threshold effect results from the finite widths of daughter
particles, which can be described by the Breit-Wigner
distribution. We illustrate the Breit-Wigner (BW) distribution
with the top quark itself in the upper plot of Fig.~\ref{tbwwid}.
We shall subsequently use an artificial distinction of whether a
particle is real or virtual: If the available energy is lower than
the central mass value by $3\,\Gamma$, we consider it as a virtual
particle in the decay final state. On the other hand, if the
available energy is more than the central mass value by $3
\,\Gamma$, it is considered as a real particle. This is indicated
as the vertical dashed band in Fig.~\ref{tbwwid}.

Let us use the simpler case of the decay width for the top
quark~\cite{Bigi:1986jk} to illustrate threshold effects involving
unstable daughter particles. In the lower part of
Fig.~\ref{tbwwid}, the decay width of the top quark, assuming
100\% branching fraction into a bottom quark and a $W$-boson, is
calculated with the convolution method~\cite{Calderon:2001qq},
treating the $W$ boson as a BW-distribution. If $m_t < m_W + m_b$,
the top quark does not have enough rest energy to produce a real
$W$ boson. As described by the Breit-Wigner distribution, this means
that the top quark mass does not fully cover the distribution of the
$W$ boson.
Once $m_t > m_W + m_b$, then most of the $W$ boson distribution
gets included. We therefore have a threshold effect:
\begin{eqnarray*}\label{eq:ttobw}
m_t &\gtrsim & m_W + m_b \ \ \Rightarrow \ \ t \to bW ,      \\
m_t &\lesssim& m_W + m_b \ \ \Rightarrow \ \ t \to bW^* . \nonumber
\end{eqnarray*}
This threshold effect caused by the $W$ width can be seen for
$m_t$ around $m_W + m_b \sim 85$ GeV. Without the finite width of
the $W$ boson, i.e. assuming the $W$ is as stable as the $b$
quark, the top quark width would drop towards zero below the $bW$
threshold.

In the case of the fourth generation $b' \to tW$ decay, we have to
consider not only the width of the $W$ boson, but also the finite
width of the top quark. The latter was not considered in the
previous study~\cite{Arhrib:2006pm}, where the oversight can be
traced to Ref.~\cite{HS1}.
Considering $t^{(*)}W^{(*)}$ as the only final state, one expects:
\begin{eqnarray*}\label{eq:bptotw}
m_{b^\prime} &\gtrsim & m_t + m_W \ \ \Rightarrow \ \ b' \to tW\,,      \\
m_{b^\prime} &\lesssim& m_t + m_W \ \ \Rightarrow \ \ b' \to tW^*\ {\rm and}\ t^*W\,.    \\
\end{eqnarray*}

\section{\label{sec:b'totW}$b^\prime \to tW$ DECAY WIDTH\protect\\}

\begin{figure}[t]
\centering
\includegraphics[width=45mm]{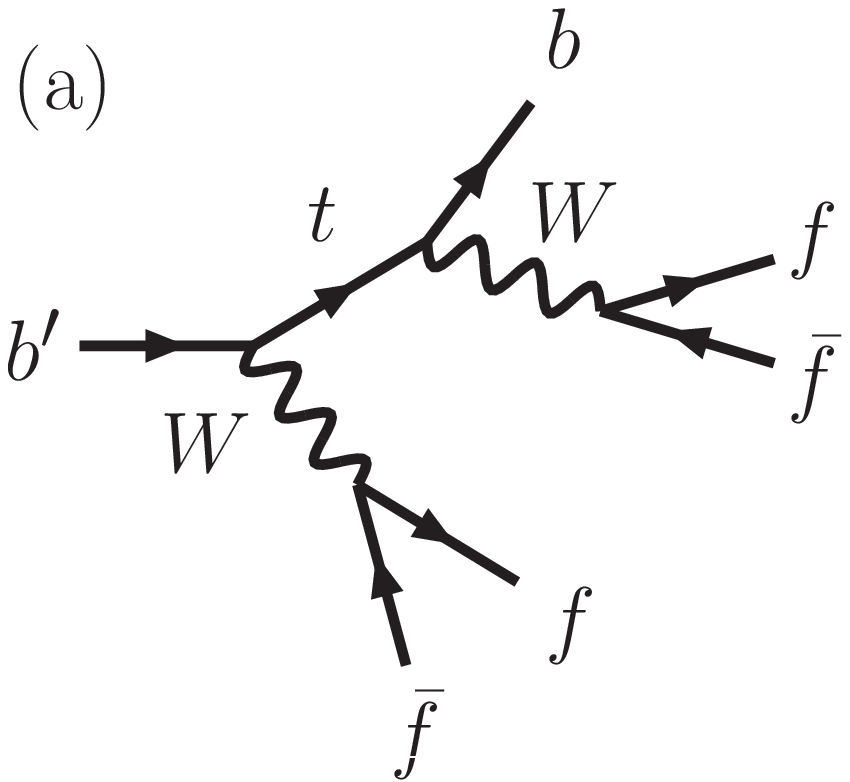}

\includegraphics[height=26mm]{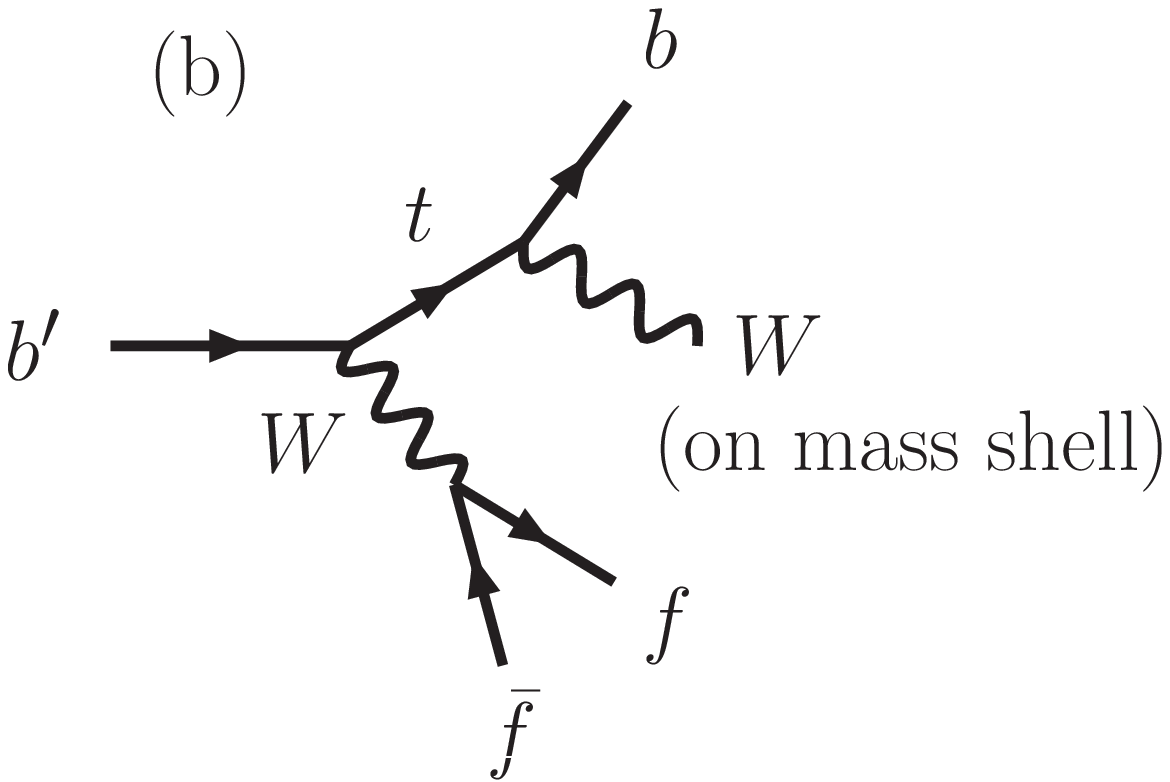}
\includegraphics[height=21mm]{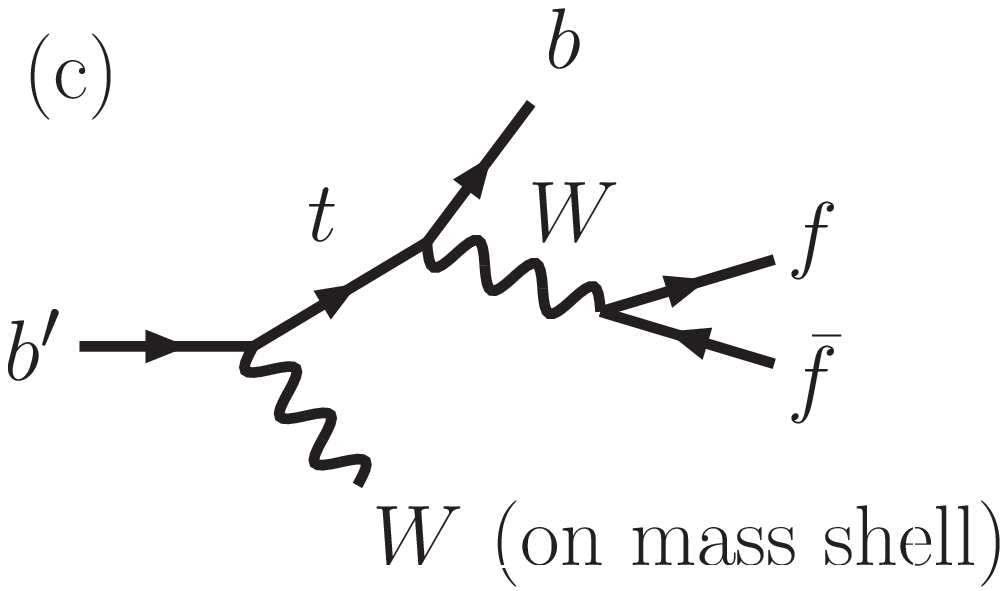}

\includegraphics[height=24mm]{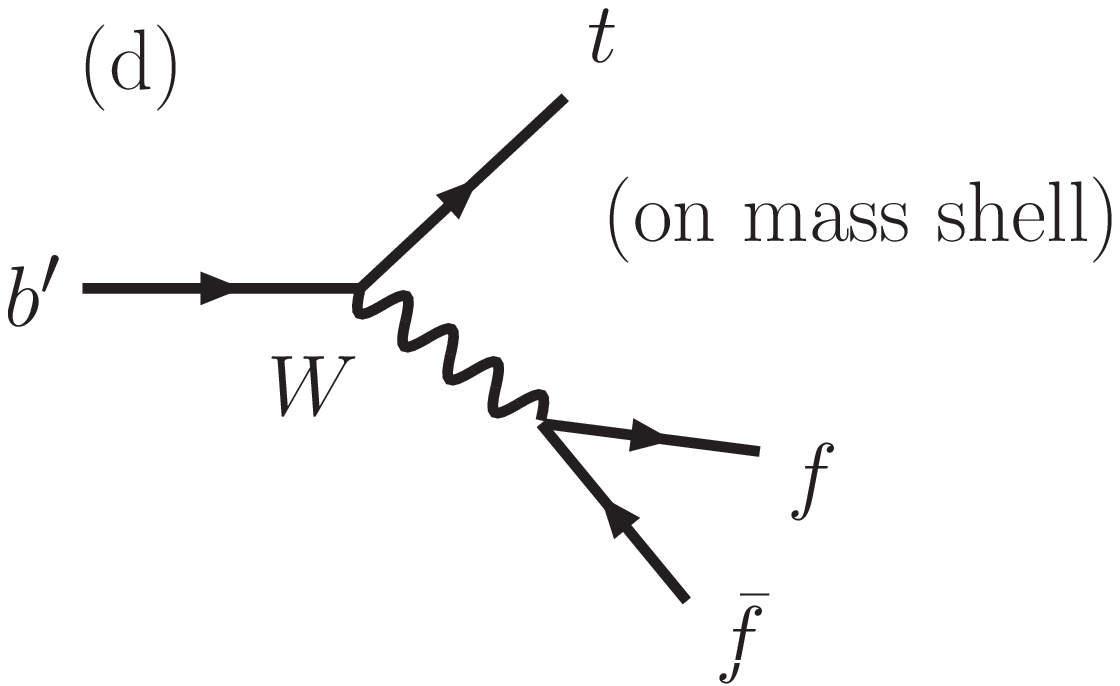}
\includegraphics[height=23mm]{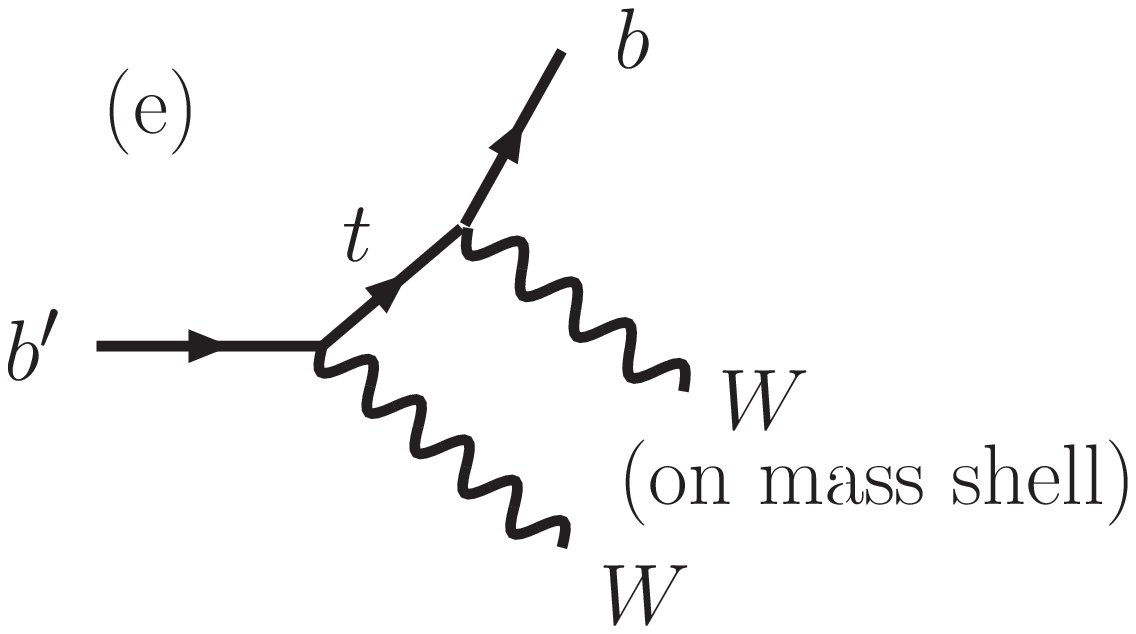}

\includegraphics[height=28mm]{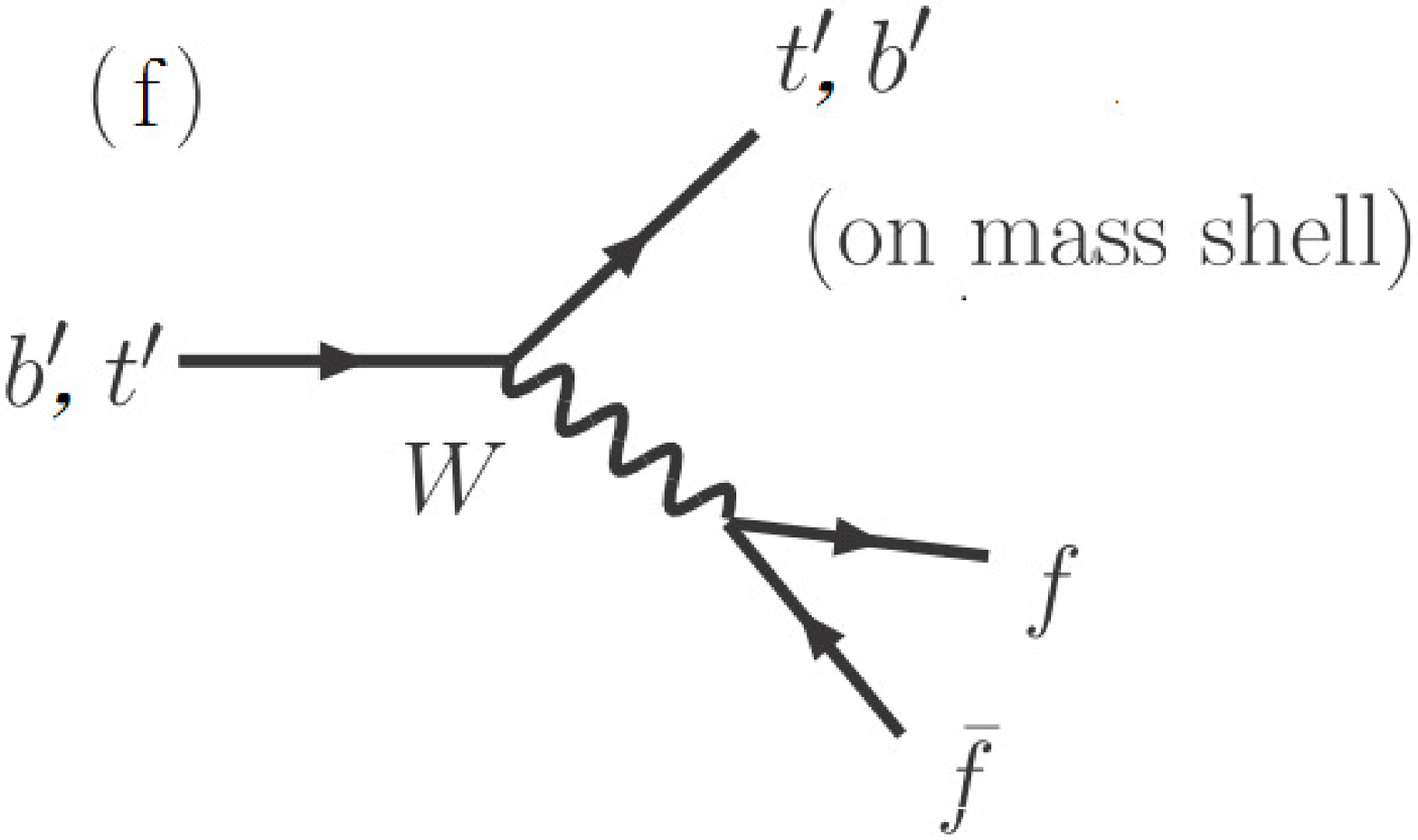}
\caption { Feynman diagrams for
 (a) five-body $b' \to t^{(*)}W^{(*)}$ decay, as well as $n$-body approximation
     diagrams under the convolution method, where
 (b) and (c) are for four-body, and
 (d) and (e) are for three-body approximations.
 The dominant three-body $b' \to t' W^*$ (or $t' \to b' W^*$)
 heavy-to-heavy transition is illustrated in (f).} \label{feynman}
\end{figure}

Let us analyze the $b' \to t^{(*)}W^{(*)}$ decay width for
different $b'$ mass values. The corresponding Feynman diagrams are
shown in Figs.~\ref{feynman}(a)--\ref{feynman}(e). By dividing the range for $m_{b'}$
into three regions via $m_{b} + 2m_{W} \lesssim m_t$ and $m_{t} +
m_{W}$ thresholds, heuristically one can evaluate the decay width
of $b'$ using the following approximations:
\begin{itemize}
  \item $m_{b'} \lesssim m_{b} + 2m_{W}$:
    $b' \to t^*W^{(*)} \to bf_i f_j W$ or $bW f_k f_l$ quasi four-body decay;

  \item $m_t \lesssim m_{b'} \lesssim m_t + m_W$:
    $b' \to t^{(*)}W^{(*)} \to bWW$ or $t f_k f_l$ quasi three-body decay;

  \item $m_{b'} \gtrsim m_t+m_W$:
    $b' \to tW$ two-body decay.
\end{itemize}

However, a direct calculation of the five-body decay width that covers
the full kinematic range for $b' \to t^{(*)}W^{(*)} \to bW^{(*)}W^{(*)}
\to b f_i f_j f_k f_l$ at tree level, can be obtained via the
convolution method~\cite{Calderon:2001qq,BarShalom:2005cf,Kuksa:2006co},
by treating $t$ and $W$ as unstable particles through a Breit-Wigner distribution.
That is
\begin{eqnarray}
\Gamma_{b' \to b f_i f_j f_k f_l}
 & = & \int^{\left(m_{b'} - m_b\right)^2}_{0}
       \int^{\left(m_{b'} - m_b - \sqrt{q^2}\right)^2}_{0} dq^2 dp^2 \nonumber \\
 && \times\; \rho(q^2, m_W, \Gamma_W^0) \, \rho(p^2, m_W, \Gamma_W^0) \nonumber\\
 && \times\; \Gamma^0(b' \to bW(q^2)W(p^2) ).
\end{eqnarray}
The three body width $\Gamma^0(b' \to bW(q^2)W(p^2))$ is for
variable effective $W$ masses $q^2$ and $p^2$, and
\begin{equation}
\rho(q^2, m_W, \Gamma_W^0) = \frac{1}{\pi}\frac{\frac{q^2}{m_W}\Gamma_W^0}
                               {\left(q^2-m_W^2\right)^2
                              + \bigl(\frac{q^2}{m_W}\Gamma_W^0\bigr)^2}
\end{equation}
is the BW distribution for the $W$ boson, where $\frac{q^2}{m_W}\Gamma_W^0$
is derived from $\sqrt{q^2}\sum_{i,j}\Gamma^0(W(p^2)\to f_i f_j)$ in
the limit of massless final state fermions~\cite{BarShalom:2005cf}.
That is, $\Gamma^0_W =\sum_{k,l} |V_{kl}|^2 \frac{N_cg_W^2m_W}{48\pi}$,
with $N_c=3$ for quarks, and $N_c=1$ for leptons. We sum over CKM dominant
$u\bar d$ and $c\bar s$ quark final states only.

The three-body width $\Gamma^0(b'\to bW(q^2)W(p^2))$ can be put as
a convolution of two-body decays,
\begin{eqnarray}
 & & \Gamma^0(b' \to bW(q^2)W(p^2)) \nonumber \\
 &=& \int_0^{\left(m_{b'} - \sqrt{p^2}\right)^2} dk^2 \; \Gamma(b' \to t(k^2)W(p^2)) \nonumber \\
 & & \times\; \rho(k^2, m_t) \, \Gamma(t(k^2) \to b(s^2)W(q^2)),
\end{eqnarray}
where $s^2 = m_b^2$, and the two-body decay width is
%
%\begin{eqnarray*}
%\Gamma^0(b' \to bW(q^2)W(p^2)) = \int^{(m_{b'}-m_b)^2}_0 \\
%\int^{(m_{b'}-m_b)^2}_0
%\int^{(m_{b'} - \sqrt{p^2})^2}_{(m_{b} + \sqrt{q^2})^2} dq^2dp^2dx^2 \\
%\frac{|V^*_{tb'}|^2|V_{tb}|^2}{2^{10}\pi^3m_{b'}m^4_W}  \frac{g^4_W}{\left[ (x^2-m^2_t)^2+(\frac{x^2}{m_t}\Gamma^0_t)^2\right]} \times \\
%\left[b^4 - 2 q^4 + q^2 x^2 + x^4 + b^2 (q^2 - 2 x^2)\right] \times \\
%\left [m^4_{b'} - 2p^4 + p^2 x^2 + x^4 + b^2 (p^2 - 2 x^2) \right] \times \\
%\sqrt{\frac{b^4 + (q^2 - x^2)^2 - 2b^2(q^2 + x^2)}{x^4}} \times \\
%\sqrt{\frac{m_{b'}^4 + (p^2 - x^2)^2 - 2m_{b'}^2(p^2 + x^2)}{m_{b'}^4}} \\
%\end{eqnarray*}
%
%\center
%
%\begin{center}
%\begin{eqnarray*}
%\Gamma^0(b' \to bW(q^2)W(p^2)) = \int\int\int dq^2dp^2dx^2 \times\\
%\frac{|V^*_{tb'}|^2|V_{tb}|^2}{2^{10}\pi^3m_{b'}m^4_W}  \frac{g^4_W}{\left[ (x^%2-m^2_t)^2+(\frac{x^2}{m_t}\Gamma^0_t)^2\right]} \times \\
%\left[b^4 - 2 q^4 + q^2 x^2 + x^4 + b^2 (q^2 - 2 x^2)\right] \times \\
%\left [m^4_{b'} - 2p^4 + p^2 x^2 + x^4 + b^2 (p^2 - 2 x^2) \right] \times \\
%\sqrt{\frac{b^4 + (q^2 - x^2)^2 - 2b^2(q^2 + x^2)}{x^4}} \times \\
%\sqrt{\frac{m_{b'}^4 + (p^2 - x^2)^2 - 2m_{b'}^2(p^2 + x^2)}{m_{b'}^4}} \\
%\end{eqnarray*}
%\end{center}
%
\begin{eqnarray}
 & & \Gamma(t(k^2) \to b(s^2)W(q^2))  \nonumber \\
 &=& \frac{G_F k^{3/2}}{8\pi\sqrt{2}}\,|V_{tb}|^2 \;
     \lambda^{1/2} \left(1,\frac{s^2}{k^2},\frac{q^2}{k^2} \right) \nonumber \\
 & & \times\; \left[ \left(1-\frac{s^2}{k^2}\right)^2
                   + \left(1+\frac{s^2}{k^2}\right)\frac{q^2}{k^2}
                   - 2\,\frac{q^4}{k^4} \right ],
\end{eqnarray}
with $\lambda(x,y,z) \equiv x^2+y^2+z^2 -2(xy+yz+xz)$, and
$\Gamma^0(b' \to t(k^2)W(p^2))$ is analogous.

\begin{figure}[t]
\centering
\includegraphics[width=65mm]{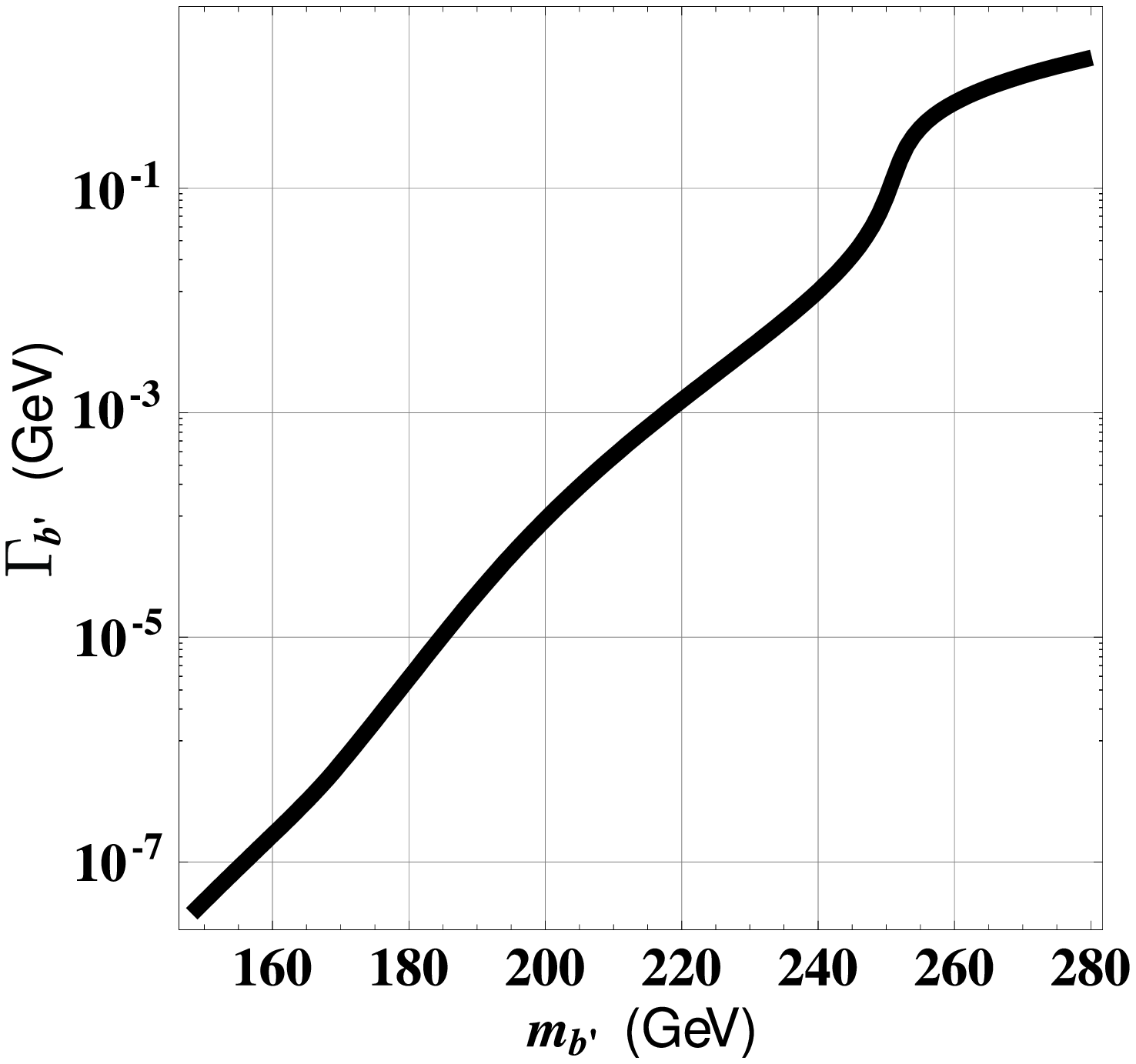}
\includegraphics[width=65mm]{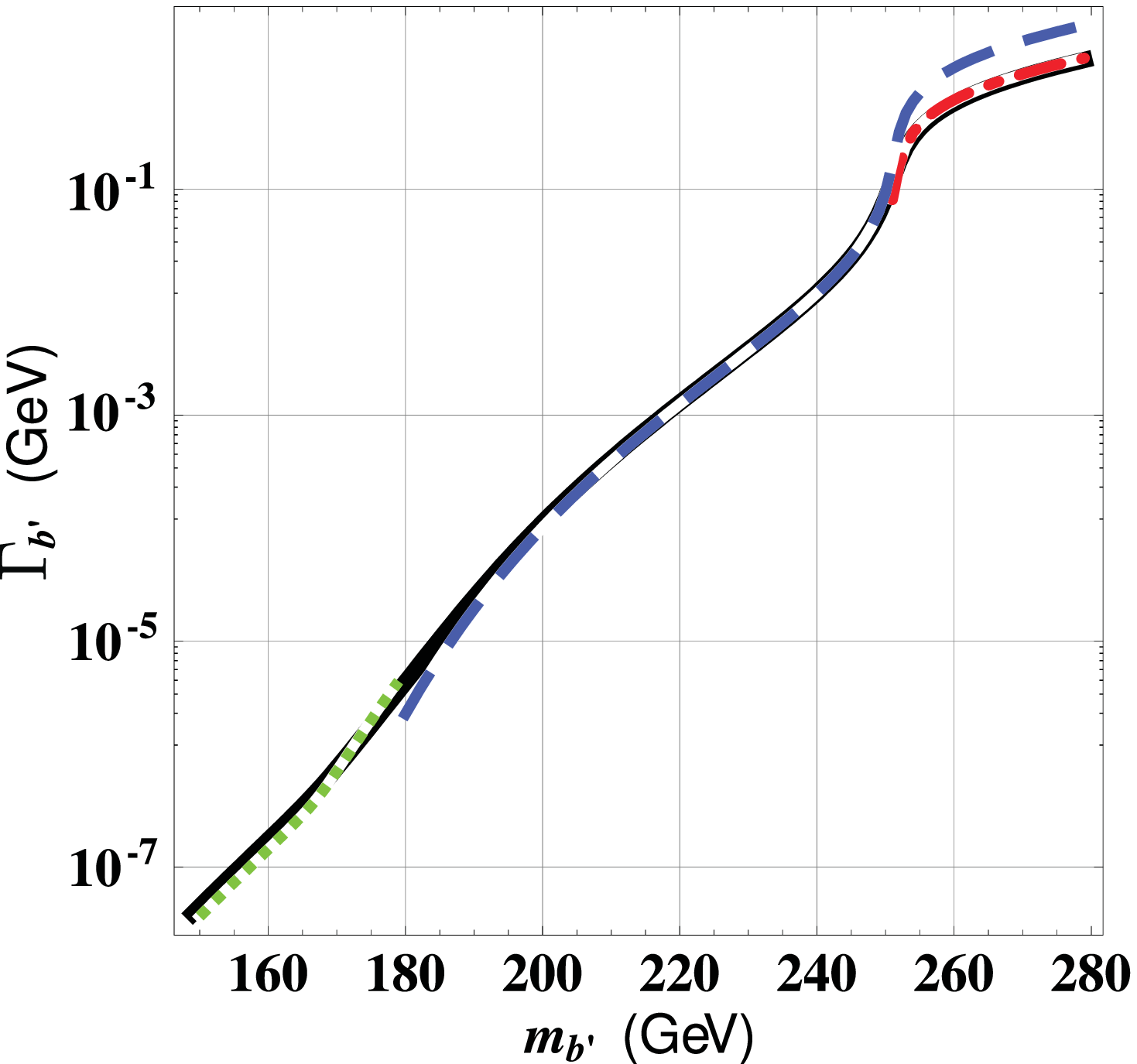}
\caption { The decay width of $b^\prime \to tW$, modulo $\vert
V_{tb'}\vert^2$. The black solid curve is the full five-body decay
result. The green dotted, blue dashed and red dot-dashed curves
show the four-, three- and two-body decay approximations, valid
for different kinematic regions. Note that extending the
three-body curve above the $tW$ threshold would result in
double-counting.
 } \label{bptwwid}
\end{figure}

After rearranging the function, using the probability distribution,
for a top quark in this case,
\begin{eqnarray}
\rho_t(q^2,m_t)=\frac{1}{\pi}
\frac{q\Gamma_t(q^2)}{(q^2-m_t^2)^2+(q\Gamma_t(q^2))^2},
\end{eqnarray}
%
%\begin{eqnarray*}
%\frac{|V^*_{tb'}|^2|V_{tb}|^2}{2^{10}\pi^3m_{b'}m^4_W}  \frac{g^4_W}{\left[ (x^%2-m^2_t)^2+(\frac{x^2}{m_t}\Gamma^0_t)^2\right]} \times \\
%\left[b^4 - 2 q^4 + q^2 x^2 + x^4 + b^2 (q^2 - 2 x^2)\right] \times \\
%\left [m^4_{b'} - 2p^4 + p^2 x^2 + x^4 + b^2 (p^2 - 2 x^2) \right] \times \\
%\sqrt{\frac{b^4 + (q^2 - x^2)^2 - 2b^2(q^2 + x^2)}{x^4}} \times \\
%\sqrt{\frac{m_{b'}^4 + (p^2 - x^2)^2 - 2m_{b'}^2(p^2 + x^2)}{m_{b'}^4}} \\
%\end{eqnarray*}
%
inserting Eq.~(2)--(5) into Eq. (1), we get the decay width of
$b'$, where the numerical result is given in Fig.~\ref{bptwwid} as
the black solid curve.
Note that Eq.~(5) is the general form of the probability distribution
for unstable particles~\cite{Kuksa:2006co},
but $q\Gamma_W(q^2)=\frac{q^2}{m_W}\Gamma_W^0$ in the limit of vanishing
final state fermion masses for $W \to f_i f_j$.
Therefore, the analogue of Eq.~(5) for the W boson reduces to Eq.~(2)
(for further discussion, see Ref.~\cite{BarShalom:2005cf}).

To make contact with various threshold effects, note that just
above a kinematic threshold, the narrow width assumption (BW
distribution becomes a $\delta$ function) can be used for some
daughter particle. Thus, in different $b'$ mass regions, we can
take the finite width effects into account and deal with the $b'$
decay processes as a $n$-body decay, with $n < 5$. In this
way, we recover the heuristic view as depicted in
Figs.~\ref{feynman}(a)--\ref{feynman}(e). The calculation is simpler, as there are fewer
phase space integrals.

In the following, we compare the full five-body decay with the
fewer body decay scenarios (see lower part of Fig.~\ref{bptwwid}).
%This approximation can greatly reduce the computation, and
%could be used for future Monte Carlo event generation.
By dividing the $m_{b'}$ into the three regions with $m_{b} +
2m_{W} \lesssim m_t$ and $m_{t} + m_{W}$ thresholds as mentioned
earlier, we can evaluate the $b'$ width using the following
approximations:

%The result of the calculation is shown as the solid black curve in
%Figure~\ref{bptwwid}. Currently, no QCD
%corrections are applied in this calculation. With the narrow-width
%approximation, one can calculate the $b'$ decay processes as n-body decays,
%where n $<5$. The calculation is simplified as one do not need to
%integrate over the full phase space.
%%The corresponding Feynman diagrams are shown in Figure~\ref{feynman}.

\begin{itemize}
\item For $m_{b'} \gtrsim m_t+m_W \sim 255$ GeV, one has an effective
two-body decay.
\item For 180 GeV $\lesssim m_{b'} \lesssim 245$ GeV, since it is
$3\,\Gamma_t$ away from $m_t$ as well as $m_t+m_W$ thresholds,
either $t$ or $W$ must be decaying off-shell. Hence, the $b'$
decay width can be estimated with a quasi-three-body decay model,
with contributions from mainly $b' \to t^* W \to bW W$ and $b' \to
tW^* \to t f_jf_k$, added incoherently. We can see from
Fig.~\ref{bptwwid} that this three-body model serves quite well
within this region, as compared with the full five-body result.
However, if one extends this approximation to $m_{b'} \gtrsim 250$
GeV, $t$ and $W$ will be both turning on-shell, such that $b' \to
bW W$ and $b' \to t f_jf_k$ become equivalent. This would give an
over-estimate of $\Gamma_{b'}$ by a factor of two when comparing
with the five-body (or $tW$ two-body) calculation, because of the
incoherent sum assumption. The two decay ``branches" are merging,
and there should be some interesting interference effects, which
would require a full five-body calculation to uncover.
\item For $m_{b'} \lesssim 160$ GeV, i.e. about $3\,\Gamma_W$ away
from $m_b + 2M_W$, the $t$ has to decay off-shell, but only one
$W$ boson can be on-shell. The effective four-body $b' \to bWf_1
f_2$ decay approximation gives consistent results with the full
five-body decay.
%
%In this region, the contribution of $b'$ decay width for $b' \to
%tW$ would be smaller than $b' \to cW$ or the FCNC channel, we will
%discuss about this in a systematical scheme.
%
\end{itemize}

\begin{figure}[t]
%\centering
%\includegraphics[width=60mm,height=40mm]{plots/tbwwid.eps}
%\vspace{30mm}
%\includegraphics[width=75mm]{plots/bpdecay.eps}
%\vspace{10mm}
\begin{minipage}{75mm}
\includegraphics[width=60mm]{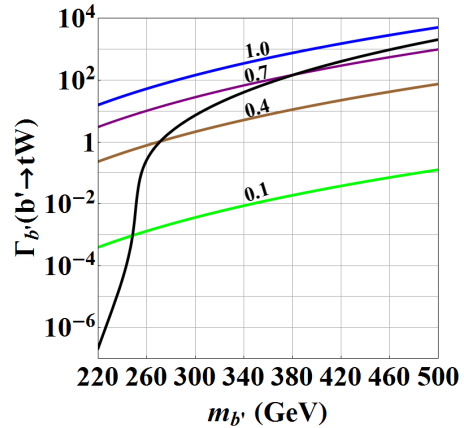}
\end{minipage}
%\hspace{-33mm}\begin{minipage}{33mm} \vspace{25mm}
%\includegraphics[width=27mm]{vcBvtB.eps}
%\end{minipage}
\caption{The decay width of $b^\prime \to tW$ (modulo $\vert
V_{tb'}\vert^2$) and $b^\prime \to cW$ for different $b'$ mass
assumptions. The $b^\prime \to tW$ curve rises from the lower left,
while the four flatter $b^\prime \to cW$ curves are for
$|V_{cb'}|/|V_{tb'}| = $ 0.1, 0.4, 0.7, 1, respectively.
%{\color{red}\it FixMe!!! Need legends. Add FCNC, too?}
 }
\label{vcbprime}
\end{figure}

\section{\label{sec:QuarkMixing}Measuring Quark Mixing via $b^\prime$ AND $t^\prime$ DECAYs\protect\\}

Different assumptions on the magnitudes of the CKM mixing elements
will lead to different dominant $t'$ and $b'$ decay channels.
The tree diagram processes $t^\prime \to bW$ and $b^
\prime \to tW$ have been treated in experimental searches so far
as the dominant decays. But even the loop suppressed FCNC decays
through penguin diagrams~\cite{Arhrib:2006pm}, $t^\prime \to tZ$, $cZ$
and $b^\prime \to bZ$, $sZ$ (even with the $Z$ replaced by $g$ and $\gamma$)
might be significant in certain kinematic or quark mixing
parameter regions.

One would naively expect the decay branching ratios of
$t'\to sW$ and $b'\to cW$ to be relatively small, because of
the jump over two generations. But we should stress that
$V_{t's}$ and $V_{cb'}$ are yet unmeasured CKM elements, and
could be unexpectedly large.
We illustrate this in Fig.~\ref{vcbprime}:
 for $m_{b'}$ considerably above $tW$ threshold,
 $b^\prime \to cW$ would dominate over $b^\prime \to tW$
 if $|V_{cb'}|/|V_{tb'}|$ is larger than 0.7.
If $|V_{cb'}|/|V_{tb'}|$ is larger than one, $b'\to cW$ will always be
larger than $b'\to tW$ because of phase space.
It would therefore be important for the experiments to separate
$t' \to bW$ and $t' \to qW$ (where $q = s,\ d$), as CDF has just
started doing, as well as separate $b'\to cW$ (even $b' \to uW$)
from the above two processes while pursuing $b'\to tW$.
If a fourth generation is discovered, we would be just at the
beginning of measuring relevant CKM elements, as in the early B physics program.

\begin{figure*}[th]
\centering
\includegraphics[width=65mm]{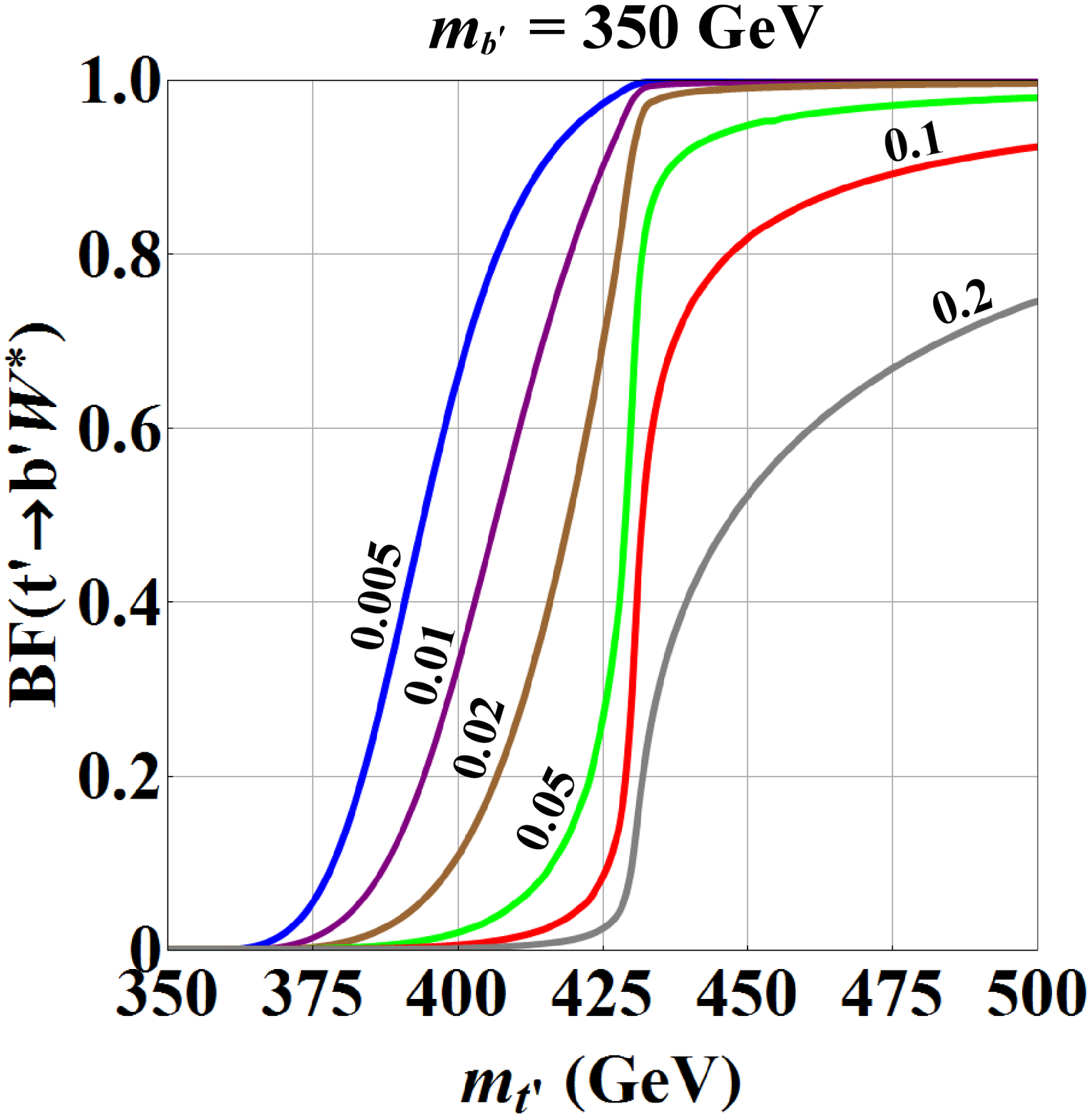}
\includegraphics[width=65mm]{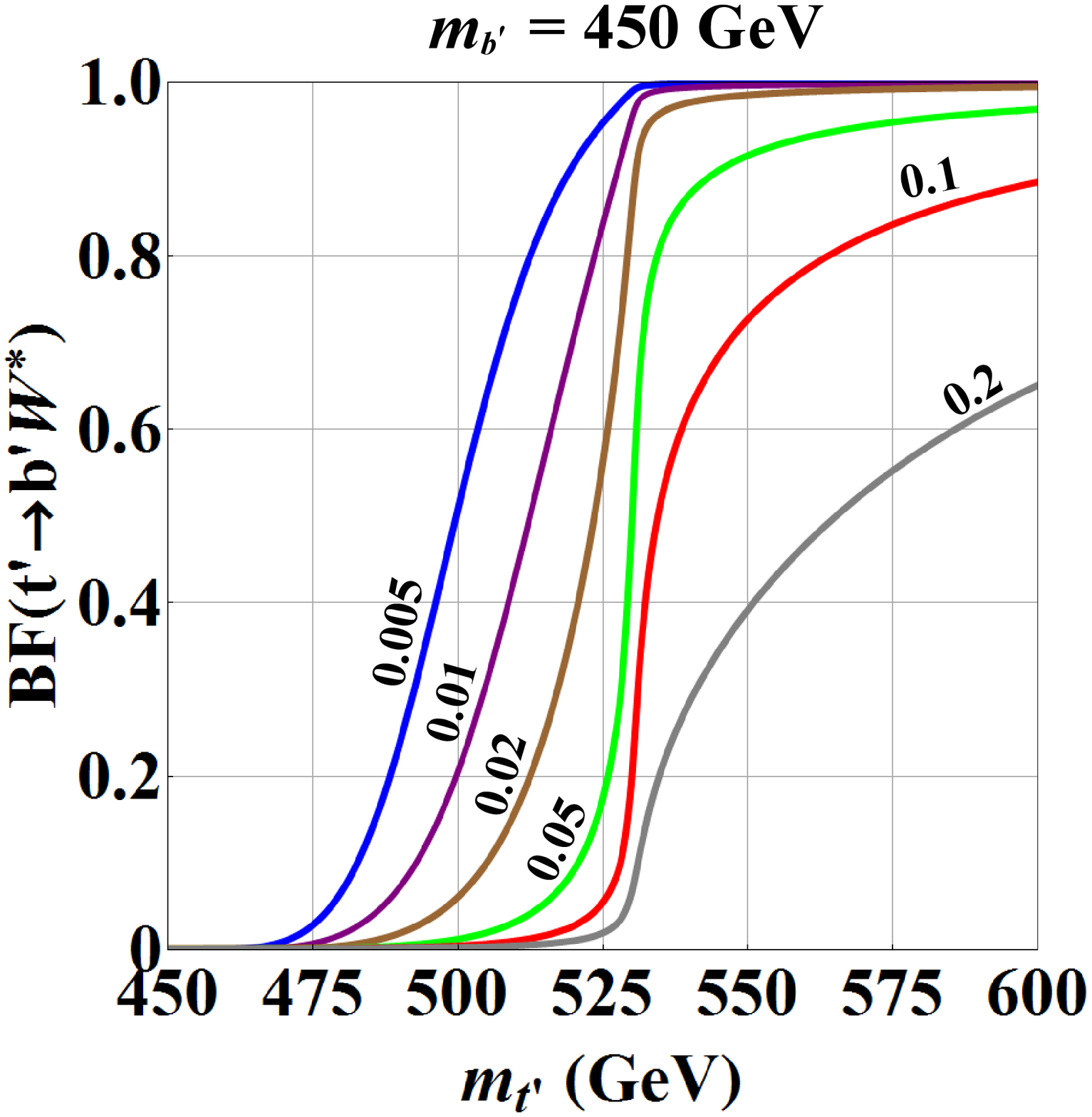}
\includegraphics[width=65mm]{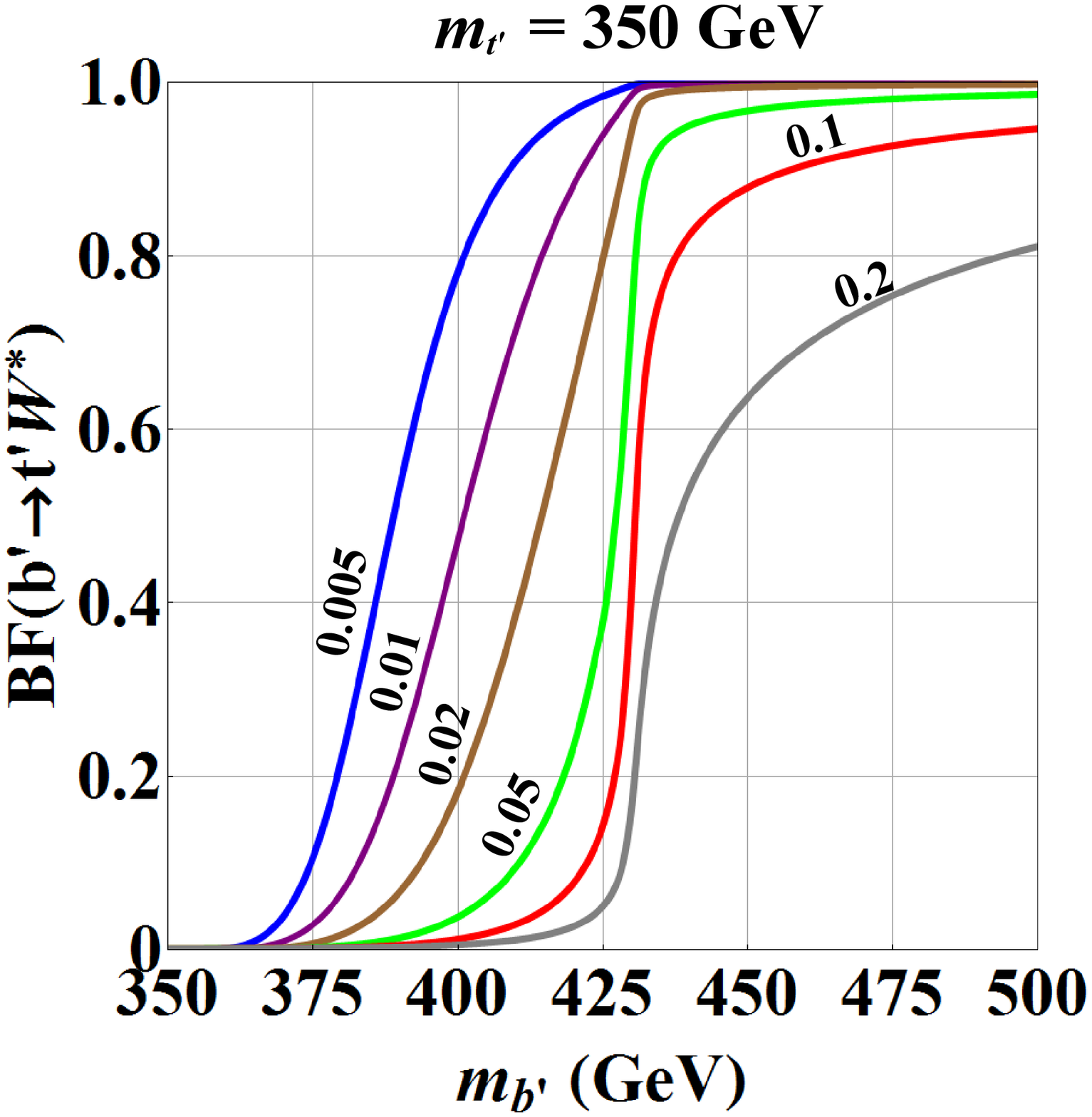}
\includegraphics[width=65mm]{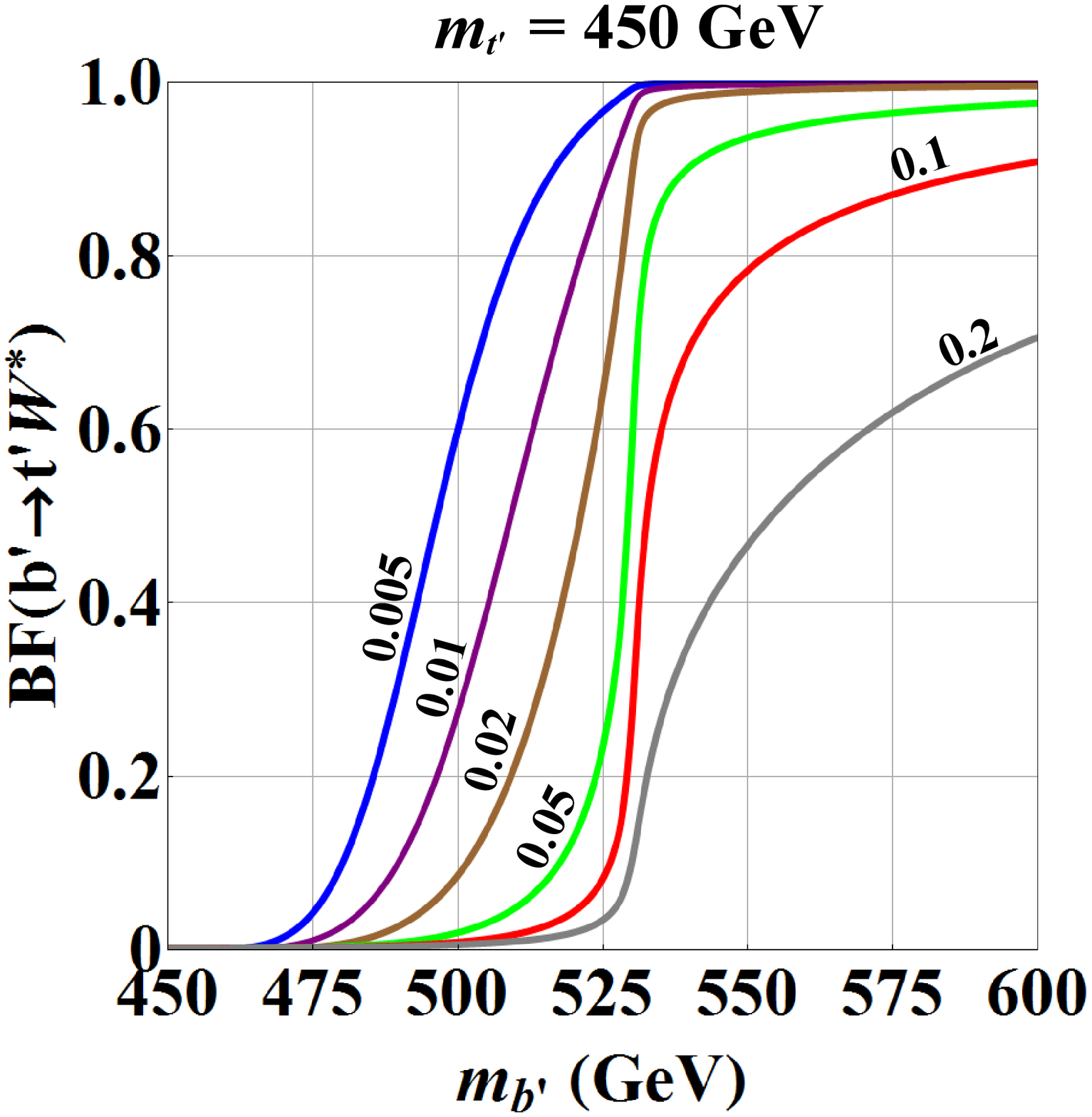}
\caption{Branching fractions for $t' \to b'W$ (upper) and $b' \to
tW$ (lower) decays. The different curves represent different
magnitudes of $|V_{t'b}|$ and $|V_{tb'}|$.
These curves would be modified if $t' \to qW$ ($q = s,\ d$) and
$b'\to cW$ are significant.
 } \label{BR}
\end{figure*}

It should be noted that, if one takes the possible hint for
sizable $t'$ effect in $b\to s$ transitions ($B \to K\pi$ direct
CPV difference, and mixing-dependent CPV in $B_s\to J/\psi\phi$)
seriously~\cite{HNS,HNS2}, then $V_{cb'}$ (related to $V_{t's}$ that
enters $b\to s$) could be comparable to $V_{tb'}$. If this is the
case, $b'\to cW$ could compete with, even dominate over $b'\to tW$
far above the $tW$ threshold!

In the same vein, it is important to keep in mind the CKM-allowed
$b'\to t'W$ decay, or $t' \to b'W$ decay, depending on whichever
quark is heavier. Electroweak precision tests (EWPrT) constrain
$|m_{t^\prime}-m_{b^\prime}| <m_W$~\cite{PDG,Kribs:2007nz,Holdom:2009rf}.
However, direct search should not be confined to the parameter space
allowed by EWPrT. Since $|V_{t^\prime b^\prime}| \simeq 1$ is rather likely,
which could be much larger than $|V_{t b^\prime}|$ and $|V_{t^\prime b}|$,
we turn to compare the CKM allowed versus the CKM suppressed $t'$ and $b'$
decays. We find that the EWPrT constraint makes the
CKM allowed intra fourth generation transitions rather interesting,
precisely because of the strong threshold dependence.

%In this section, the branching fractions (BF) of $t^\prime$ and $b^\prime$
%and their relationships to the CKM matrix elements are discussed. Their
%decay widths are calculated in tree level.

First, let us consider $m_{t'} > m_{b'}$. The width
for the top-like decay $t' \to bW$ is proportional to $|V_{t'b}|^2$.
As $|V_{t'b'}| \simeq |V_{tb}| \simeq 1$, while the value of
$|V_{tb'}|^2$ is expected to be of order 1\% or less,
$t' \to b^{\prime *}W$ (where $b'$ is virtual)
is suppressed by both phase space and $|V_{tb'}|^2$.
Thus, to very good approximation, we can treat $b'$ as
on-shell in the final state (we have verified this by
direct computation of $t' \to b^{\prime *}W$).
The EWPrT bound that $m_{t'} - m_{b'} < M_W$ then implies that the associated
$W$ is off-shell, as illustrated in Fig.~\ref{feynman}(f). The three-body
phase space suppression could be compensated by the CKM allowed coupling,
as compared with the two-body phase space but CKM suppressed $t' \to bW$ decay.

Assuming $t' \to b'W^*$ and $t' \to bW$ to be the two dominant modes,
we plot the branching fraction (BF) of $t' \to b'W^*$ in the upper two plots
of Fig.~\ref{BR} as a function of $m_{t^\prime}$, for
$m_{b^\prime} =$ 350 and 450 GeV, and for
$|V_{t'b}| =$ 0.005, 0.01, 0.02, 0.05, 0.1, and 0.2.
For $m_{b'}$ just above $m_{t'}$, the BF for $t' \to b'W^*$ is severely suppressed.
However, depending on how small $|V_{t'b}|$ is, hence how much $t'\to bW$
rate gets suppressed, the BF for $t' \to b'W^*$ can become considerably enhanced.
So, if one knows $b'$ mass already, then \emph{a measurement of BF($t' \to b'W^*$)
would provide a measurement of $|V_{t'b}|$}. It is particularly sensitive
to small $|V_{t'b}|$ values (this sensitivity drops somewhat as the $t'$, $b'$
system becomes heavier), which provides a complementary program to
the indirect measurement through loop-induced $b\to s$ transitions.
If it so happens that the EWPrT constraint of $m_{t'} - m_{b'} < M_W$ is violated,
hence the $W$ in $t' \to b'W$ decay is on-shell, then this heavy-to-heavy process
would in fact dominate for small $|V_{t'b}|$ values. One should then measure
BF($t' \to bW$) instead for the determination of $|V_{t'b}|$.

%
%Although $t' \to b'W$ is CKM-favored because of only the same generation
%quark involved, it is limited by the phase space constraint from $m_{t'}$
%and $m_{b'}$ mass difference. On the contrary, $t' \to bW$ is not
%constrained by the phase space, but it is suppressed by a smaller CKM
%matrix element $|V_{t'b}|$ amplitude due to quark generation crossing.

The case for $m_{b'} > m_{t'}$ is analogous. The decay width of $b' \to
tW$ is proportional to $|V_{tb'}|^2$, which is almost the same as $|V_{t'b}|^2$
and not more than 1\%, while $|V_{t'b'}| \simeq~1$.
We give BF($b' \to t'W^*$) in the lower two plots of Fig.~\ref{BR}
as a function of $m_{b'}$, for $m_{t^\prime} =$ 350, 450 GeV,
and for $|V_{tb'}|=$ 0.005, 0.01, 0.02, 0.05, 0.1, 0.2.
When the EWPrT constraint is violated (i.e. $m_{b'} > m_{t'} + m_{W}$),
$b' \to t'W$ decay could dominate. But for $m_{b'} -  m_{t'} < M_W$,
a measurement of BF($b' \to t'W^*$) provides a measure of $|V_{tb'}|$.
Because $b' \to tW$ decay itself is kinematically limited
compared to $t' \to bW$ decay,
BF($b' \to t'W^*$) is slightly more sensitive to $|V_{tb'}|$
than BF($t' \to b'W$) is to $|V_{t'b}|$.

As we have mentioned the potential importance of $b' \to cW$ and $t' \to sW$
transitions as compared to $b' \to tW$ and $t' \to bW$, we remind the
fourth generation searchers at colliders that the true target is quite broad:
either $t' \to sW,\ bW$ and $b'W^*$, and $b' \to cW,\ tW$;
 or $t' \to sW,\ bW$, and $b' \to cW,\ tW$ and $t'W^*$
(there can still be $t' \to dW$ and $b' \to uW$).
In the first case where $t'$ is heavier, $b'$ would likely be first
discovered via $b' \to cW$ and especially $b' \to tW$.
But one has to not only separate $t' \to sW$, $t' \to bW$ and $b'\to cW$
(which all feed the current $Q \to qW$ search signature~\cite{Whiteson10}),
but also disentangle the complicated $t' \to b'W^*$, which, if dominant,
would involve a $t'\bar t' \to t\bar tW^+W^-W^{*+}W^{*-} \to
b\bar bW^+W^-W^+W^-W^{*+}W^{*-}$ final state.
In the second case where $b'$ is heavier, one should first discover
a new ``heavy top" quark, then try to separate $t' \to sW$, $t' \to bW$
and $b'\to cW$, while disentangling $b' \to tW$ (with $t \to bW$)
from $b'\to t'W^*$ (with $t'\to sW$, $bW$). That is, for $b'$ production,
besides $b' \to cW$ possibility which becomes part of the $t'$ program,
in the same-sign dilepton approach arising from $q_1\bar q_2WWW^{(*)}W^{(*)}$,
the signature is potentially rather complex, where there could be anywhere
from zero to two $b$-tagged jets, while one or two $W$ boson could be off-shell.

\section{\label{sec:Conclusion} Discussion and Conclusion\protect\\}

Our starting point was noting that, in the case of the
fourth generation $b' \to tW$ decay, not only the width of the $W$ boson,
but also the width of the top quark have to be considered.
A direct calculation of the five-body decay width, obtained via
the convolution method, can cover the full kinematic range for
$b' \to t^{(*)}W^{(*)} \to bW^{(*)}W^{(*)} \to b f_i f_j f_k f_l$
at tree level. One can also check the various effective four-,
three- and two-body decay processes.
In so doing, we clarified how the two branches of
$b' \to t^*W$ and $b' \to tW^*$ merge to $b' \to tW$, where both
$t$ and $W$ are on-shell.
This is, in fact already incorporated in PYTHIA.
We compare our results with that of PYTHIA~6 in the Appendix.
Though the general trend is consistent, some difference is noticed.
We note that a full five-body calculation is needed to uncover the
interesting interference effects which happen while the above-mentioned
two ``branches" are merging.

Our computation, though easily adapted to the $b' \to t'W^*$ (or $t' \to b'W^*$)
case, certainly did not consider initial and final state interactions.
For these, and to make experimental contact, one would need to link with
the pair production, as well as jet fragmentation processes.
One would then need to incorporate various QCD corrections, which
is certainly outside the scope of this work. Note that the case
of $b' \to t'W^*$ (or $t' \to b'W^*$) would be in a somewhat different
kinematic regime than $b' \to tW$ for these correction, given that
$b'$ and $t'$ are semi-degenerate, especially when their mass scale becomes higher.

The dominant decay channels of $t'$ and $b'$ quarks depend not only on
their masses and mass difference, but on the CKM mixing elements,
$V_{t'b}$, $V_{t's}$ (and $V_{t'd}$), or $V_{tb'}$, $V_{cb'}$ (and $V_{ub'}$)
as well. We have illustrated with the two cases of
suppressed $t' \leftrightarrow b'$ transitions (Fig.~\ref{vcbprime}),
or suppressed $b' \to cW$ and $t' \to sW$ decays (Fig.~\ref{BR}).
In the former case, separating say $t'\to sW$ from $t' \to bW$
(measuring BF($t'\to sW$))
could provide a measurement of $|V_{t's}/V_{t'b}|$, while
in the latter case, a measurement of BF($t' \to b'W^*$)
would provide a measurement of $|V_{t'b}|$, with the sensitivity
geared towards small $|V_{t'b}|$ values.
The complete program, alluded to at the end of the previous section,
involving fourth generation quark decays to fourth, third, second
and even first generation quarks, is much more complex.

Let us just consider the case of ``classical splitting"~\cite{Whiteson10},
i.e. $t'$ heavier than $b'$, and satisfying the EWPrT constraint
of $m_{t'} < m_{b'} + M_W$. One should consider the decays
$b' \to tW,\ cW,\ uW$, and $t' \to b'W^*,\ bW,\ sW,\ dW$.
For simplicity, let us drop the first generation. Ref.~\cite{Whiteson10}
discussed how the various channels feed the two current search channels
of lepton plus multijets and missing transverse momentum,
and same-sign dileptons plus jets and missing transverse momentum.
Their main point is that the limits would be in general improved
by combining the two studies.
But they noted that if BF($t' \to b'W$) is sizable,
it would weaken the $b'$ mass bound.

\begin{figure*}[t]
\centering
\includegraphics[width=85mm]{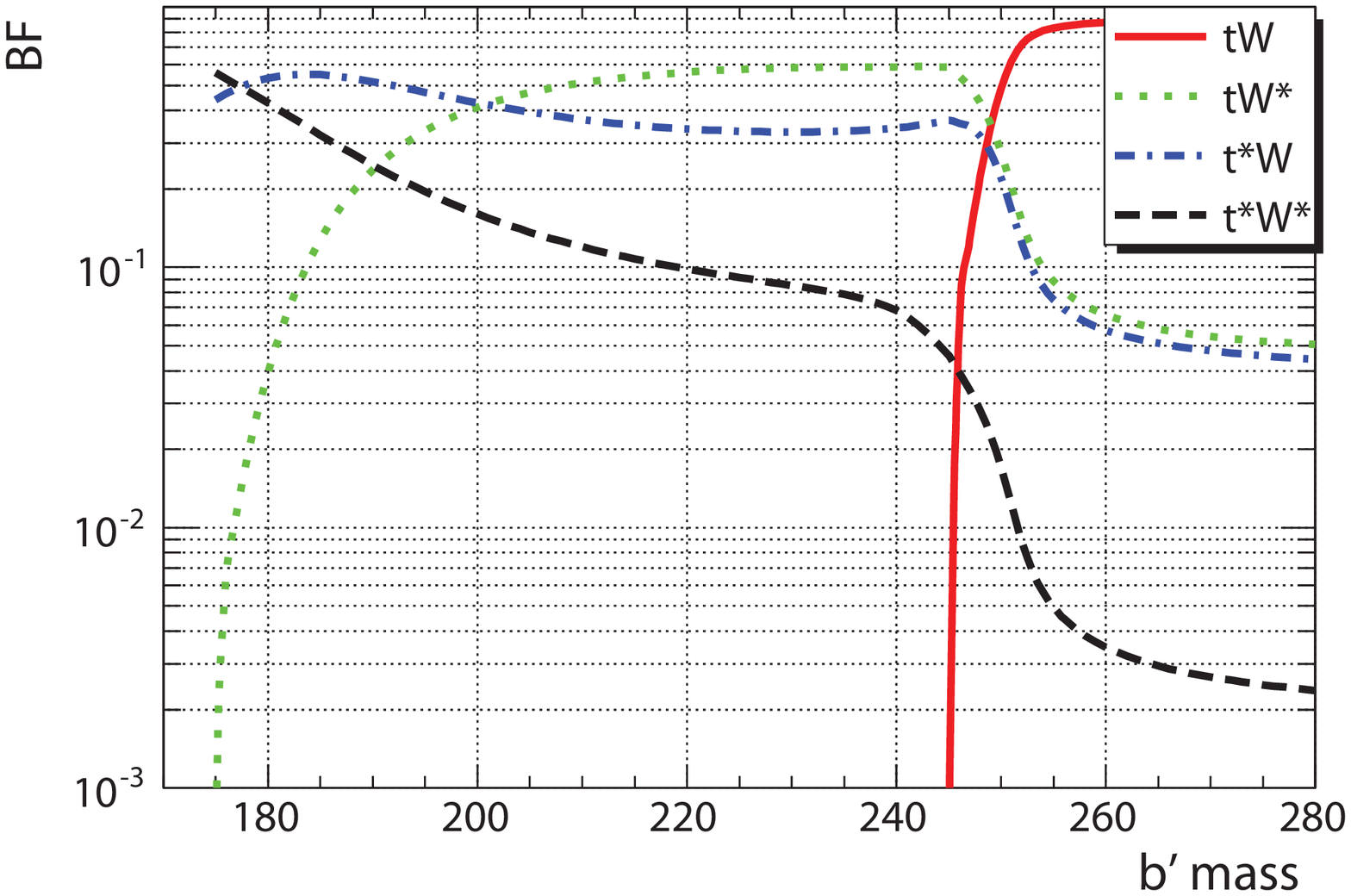}
\includegraphics[width=85mm]{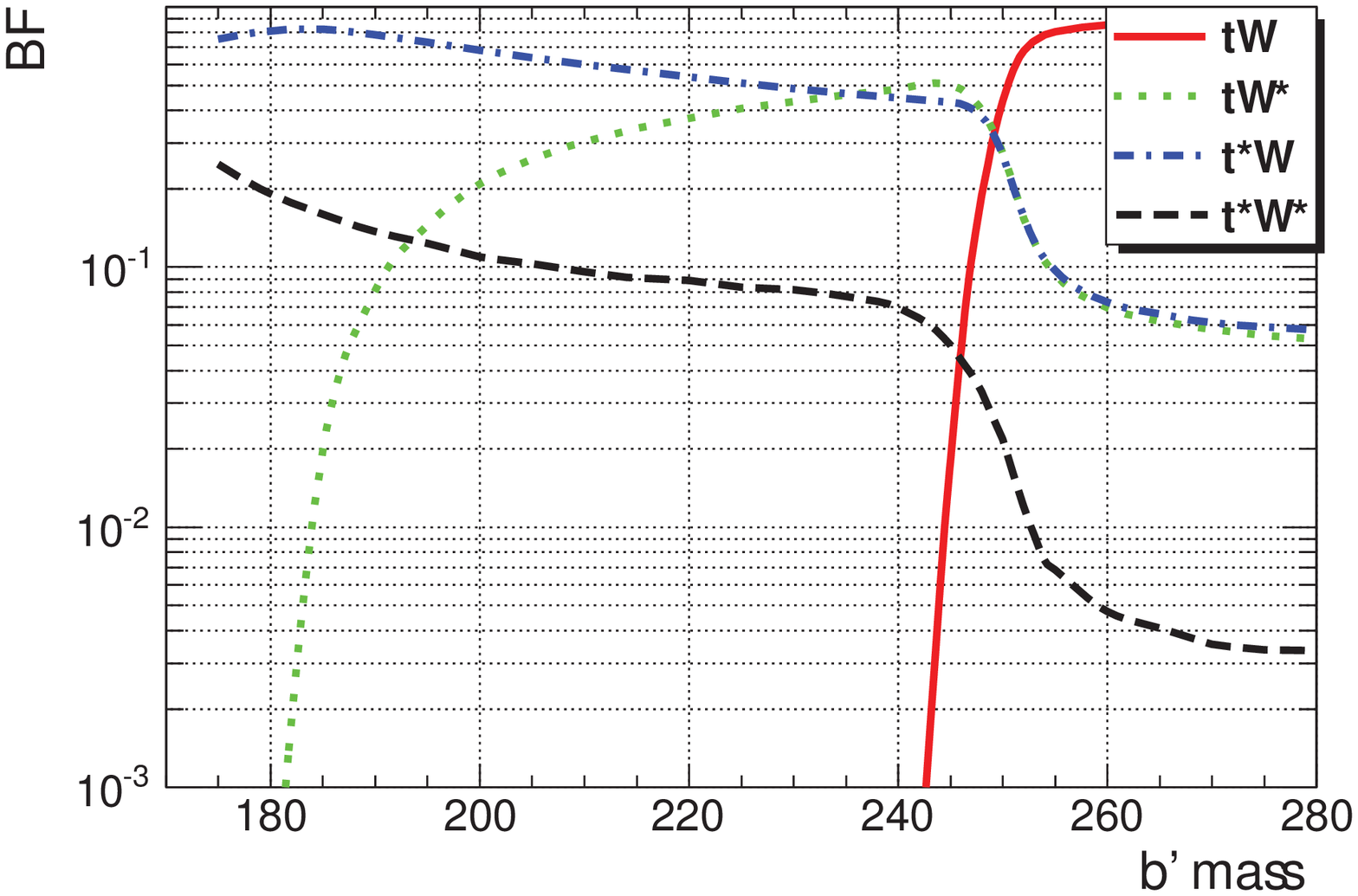}
\caption {
The branching fractions for $b' \to tW$, $tW^*$, $t^*W$,
and $t^*W^*$ processes, corresponding to two-body, three-body
and five-body processes as separated by kinematics.
The (red) solid, (green) dotted, (blue) dot-dashed
and the (black) dashed lines are for $b' \to tW$,
$tW^*$, $t^*W$ and $t^*W^*$, respectively.
The left plot is from our five-body calculation,
while the right plot is obtained from PYTHIA 6~\cite{Sjostrand:2006za}.
The trend of the two plots are similar, but differ in the details.
 } \label{pythia}
\end{figure*}

Our point is beyond Ref.~\cite{Whiteson10}, that
there is a wealth of information on CKM quark mixing
involving the fourth generation,
if one could separate the various $t'$ and $b'$ quark decay modes.
The $b'\to tW$ mode would be the easiest to uncover via
the same-sign dilepton signature, even if $b' \to cW$ decay is sizable.
But the presence of the latter would complicate the agenda,
with further complication through the $t' \to b'W^*$ mode.
New methods would have to be developed (e.g. the boosted-$W$ technique
used by ATLAS~\cite{ATLAS1}) to disentangle such complicated events
of $t\bar tWW$, $t\bar cWW$, with the possible addition of up to
two, perhaps off-shell, $W$ bosons.
The analysis should certainly go beyond the discovery method of
same-sign dileptons, for example a fully reconstructed top plus
a boosted-$W$ imbedded in a rather complicated event.
At the same time, there would be a separate analysis track of
$c\bar cWW$, $b\bar bWW$, $b\bar sWW$ and $s\bar sWW$,
similar to the current $Q \to qW$ search program.
The $b$ (and $c$) tagging separation, already invoked by CDF~\cite{Ivanov},
must be applied, together with an effort to separate the mass
and charge of the decaying parent.
In the case of prominent $b' \to cW$, the $t'\to b'W^*$ mode
could result in $c\bar cWWW^*W^*$ events that have no trace of
top quarks in them.

A main consequence of the consideration of threshold behavior
for $t' \to b'W^*$ is that, if this decay mode can be separated
in the above analysis (and with $t'$ and $b'$ decays separated),
we have an ``amplifier" for the measurement of small $|V_{t'b}|$ values
in the measurement of BF($t' \to b'W^*$), if $m_{t'} < m_{b'} + M_W$ is satisfied.
Consideration of $t' \to sW$ implies that one really measures the
ratio of $t' \to bW$ and $t' \to b'W^*$ decay rates, while the ratio with
$t' \to sW$ would provide information on $|V_{t's}|$.
The analysis can be extended if in fact $m_{t'} > m_{b'} + M_W$ is found,
though one would then be more sensitive to modest, rather than very small
$|V_{t'b}|$ values.

In conclusion,
with the fast rise in accumulated luminosity at the LHC,
one expects great progress in the search for fourth generation
$t'$ and $b'$ quarks, with good potential for discovery.
If discovery is made, the next task would be to sort out
all decay modes. For this matter, it is important to cover,
and separate, the CKM suppressed decays $t' \to sW$ and $b' \to cW$.
Equally important would be to search for either $t' \to b'W^{(*)}$,
or $b' \to t'W^{(*)}$, where the electroweak precision test constraint
of $|m_{t'} - m_{b'}| < M_W$ would imply that the associated $W$ boson
is virtual. The decay final states of these very heavy chiral quarks
would be rather complex, but the threshold sensitivity studied in this
work suggests that a measurement of the  $t' \to b'W^{(*)}$
(or $b' \to t'W^{(*)}$) decay branching fraction, as compared with
$t' \to bW$ and $sW$ (or $b' \to tW$ and $cW$), would provide a
sensitive measurement of $|V_{t'b}|$ or the combined strength of
$V_{t'b}$ and $V_{t's}$ ($|V_{tb'}|$ or the combined strength of
$V_{tb'}$ and $V_{cb'}$). This would complement the indirect
studies of loop-induced $b \leftrightarrow s$ transitions for
an enlarged quark mixing sector.

\appendix*

\section{\label{sec:PYTHIA}A COMPARISON WITH PYTHIA\protect\\}

In Sec.~III, we compared our various calculations of the
$b^\prime \to t^{(*)}W^{(*)}$ decay width with the result
for five-body final state. Even though $m_{b^\prime} \lesssim 300$ GeV
is excluded by experimental data, it is of interest to compare the
calculated decay widths with those from the
PYTHIA 6 generator~\cite{Sjostrand:2006za}, at least as a cross-check.
The results are shown side-by-side in Fig.~\ref{pythia}.
For the PYTHIA results, we identify an on-shell or off-shell decay
with the definition described in Sec.~\ref{sec:II} of this report.
That is, by comparing the mass of the decaying $t$ or $W$ with its
central value, if the applied energy is larger than the central value
of the particle mass three times of the natural width, the decay is
identified as on-shell, as shown in Fig.~\ref{tbwwid}. The same
definition is used in the integrations of $n$-body model calculations.
Although the trends of our calculation and the PYTHIA results are similar,
we can see quantitative deviations. Clearly, the effects of
initial and final state interactions (ISR/FSR) as well as other corrections
that are built into PYTHIA are not considered in our calculation.
The actual cause of the deviations would need further investigation to
clarify.


\begin{thebibliography}{0}
\expandafter\ifx\csname natexlab\endcsname\relax\def\natexlab#1{#1}\fi
\expandafter\ifx\csname bibnamefont\endcsname\relax
  \def\bibnamefont#1{#1}\fi
\expandafter\ifx\csname bibfnamefont\endcsname\relax
  \def\bibfnamefont#1{#1}\fi
\expandafter\ifx\csname citenamefont\endcsname\relax
  \def\citenamefont#1{#1}\fi
\expandafter\ifx\csname url\endcsname\relax
  \def\url#1{\texttt{#1}}\fi
\expandafter\ifx\csname urlprefix\endcsname\relax\def\urlprefix{URL }\fi
\providecommand{\bibinfo}[2]{#2}
\providecommand{\eprint}[2][]{\url{#2}}

\end{thebibliography}


\begin{thebibliography}{99}   % Use for  1-9  references
%\begin{thebibliography}{99} % Use for 10-99 references

%
\bibitem{Kobayashi:1973fv}
  M.~Kobayashi and T.~Maskawa,
%  ``CP Violation In The Renormalizable Theory Of Weak Interaction,''
  Prog.\ Theor.\ Phys.\  {\bf 49}, 652 (1973).

%
\bibitem{PDG}
  K. Nakamura \textit{et al.} [Particle Data Group], J. Phys. G \textbf{37}, 075021 (2010).

%
\bibitem{Hou:2008xd}
  W.-S.\ Hou,
%  ``CP Violation and Baryogenesis from New Heavy Quarks,''
  Chin.\ J.\ Phys.\ {\bf 47}, 134\ (2009) [arXiv:0803.1234\ [hep-ph]].

%
\bibitem{Kribs:2007nz}
  G.D.~Kribs, T.~Plehn, M.~Spannowsky and T.M.P.~Tait,
%  ``Four generations and Higgs physics,''
  Phys.\ Rev.\  D {\bf 76}, 075016 (2007)
  [arXiv:0706.3718 [hep-ph]].

%
\bibitem{Holdom:2009rf}
  For a recent brief review on the 4th generation, see
  B.~Holdom, W.-S.~Hou, T.~Hurth, M.L.~Mangano, S.~Sultansoy and G.~\"{U}nel,
%  ``Four Statements about the Fourth Generation,''
  PMC Phys.\  A {\bf 3}, 4 (2009)
  [arXiv:0904.4698 [hep-ph]].


%\bibitem{Scott:2006}
% CDF Collaboration, A.~L. Scott and D.~Stuart,
% ``Search for New Particles Decaying to $Z^0$+jets,''
% {\em CDF Note} {\bf 8590} (2006).
%
\bibitem{CDFt} A. Lister (for the CDF collaboration),
presented at ICHEP 2010, arXiv:1105.5992v2 [hep-ex].

\bibitem{Ivanov} Talk by Andrew Ivanov,
presented at FPCP 2011.

\bibitem{D0} V.M. Abazov {\it et al.}  [D0 Collaboration],
arXiv:1104.4522 [hep-ex].

%\bibitem{aaltonen-2008-100}
% CDF Collaboration, T.~Aaltonen,
% ``Search for Heavy Top-like Quarks $t' \to Wq$ Using Lepton Plus Jets Events
%  in 1.96 TeV Proton-Antiproton Collisions,''
%  Phys. Rev. Lett. 100, 161803 (2008)
%% url = {doi:10.1103/PhysRevLett.100.161803},
%
\bibitem{Aaltonen:2009nr}
  T.~Aaltonen {\it et al.}  [CDF Collaboration],
  %``Search for New Bottomlike Quark Pair Decays $Q\Qbar \to
  %(\t\Wmp)(\tbar\Wpm)$ in Same-Charge Dilepton Events,''
  Phys.\ Rev.\ Lett.\  {\bf 104}, 091801 (2010)
  [arXiv:0912.1057 [hep-ex]].

\bibitem{Aaltonen:2011vr} T. Aaltonen \textit{et al.} (CDF Collaboration),
  %``Search for heavy bottom-like quarks decaying to an electron or muon and jets in $p\bar{p}$ collisions at $\sqrt{s}=1.96$ {TeV}'',
  Phys.\ Rev.\ Lett. {\bf 106}, 141803 (2011) [arXiv:1101.5728 [hep-ex]].
%%CITATION = 1101.5728;%%.

\bibitem{CMS1} S.~Chatrchyan {\it et~al.} [CMS Collaboration], [arXiv:1102.4746 [hep-ex]], to appear in Phys.~Lett.~B.

\bibitem{ATLAS1}
   The ATLAS Collaboration, ATLAS-CONF-2011-022.

\bibitem{Mahlon:1995co}
  G.~Mahlon and S.J.~Parke, Phys.\ Lett.\  B {\bf 347}, 394 (1995)
  [arXiv:hep-ph/9412250].

\bibitem{Altarelli:2000nt}
  G.~Altarelli, L.~Conti and V.~Lubicz,
%  ``The $t \to W Z b$ decay in the standard model: A critical reanalysis,''
  Phys.\ Lett.\  B {\bf 502}, 125 (2001)
  [arXiv:hep-ph/0010090].

\bibitem{Calderon:2001qq}
  G.~Calder\'{o}n and G.~L\'{o}pez Castro,
%  ``Finite W boson width effects in the top quark width,''
  Int.\ J.\ Mod.\ Phys.\  A {\bf 23}, 3525 (2008)  [arXiv:hep-ph/0108088].
 %%CITATION = HEP-PH/0108088;%%


\bibitem{BarShalom:2005cf}
 S.~Bar-Shalom, G.~Eilam, M.~Frank and I.~Turan,
% ``Width effects on near threshold decays of the top quark $t \to c W W$, $c Z Z$ and of neutral Higgs bosons,''
 Phys.\ Rev.\  D {\bf 72}, 055018 (2005)
 [arXiv:hep-ph/0506167].
 %%CITATION = PHRVA,D72,055018;%%

\bibitem{Kuksa:2006co}
  V.I. Kuksa,
%  ``The convolution formula for a decay rate''
  Phys.\ Lett.\  B {\bf 633}, 545 (2006) [arXiv:hep-ph/0508164v3].

%
\bibitem{Arhrib:2006pm}
  A.~Arhrib and W.-S.~Hou,
  %``Flavor changing neutral currents involving heavy quarks with four generations,''
  JHEP {\bf 0607}, 009 (2006)
  [arXiv:hep-ph/0602035].

%
\bibitem{Bigi:1986jk}
  I.I.Y.~Bigi, Y.L.~Dokshitzer, V.A.~Khoze, J.H.~K\"uhn and P.M.~Zerwas,
  %``Production and Decay Properties of Ultraheavy Quarks,''
  Phys.\ Lett.\  B {\bf 181}, 157 (1986).


%
\bibitem{HS1}
  W.-S.~Hou and R.G.~Stuart
%  `` On Discovering The Next Charge $-1/3$ Quark Through Its Flavor Changing Neutral Current Decays,''
  Phys.\ Rev.\ Lett.\  {\bf 62}, 617 (1989);
%  ``Flavor Changing Neutral Currents Involving Heavy Fermions: A General Survey,''
  Nucl.\ Phys.\  B {\bf 320}, 277 (1989).

%
\bibitem{HNS}
  W.-S.~Hou, M.~Nagashima and A.~Soddu,
%  ``Difference in $B^+$ and $B^0$ direct $CP$ asymmetry as effect of a fourth generation,''
  Phys.\ Rev.\ Lett.\  {\bf 95}, 141601 (2005)
  [arXiv:hep-ph/0503072];
\bibitem{HNS2}
%  ``Enhanced $K_L \to \pi^0 \nu \bar\nu$ from direct CP violation in $B \to K \pi$ with four generations,''
  W.-S.~Hou, M.~Nagashima and A.~Soddu, Phys.\ Rev.\  D {\bf 72}, 115007 (2005)
  [arXiv:hep-ph/0508237];
%  ``Large time-dependent $CP$ violation in $B_s^0$ system and finite $D^0$-$\bar{D}^0$ mass difference in four generation standard model,''
  {\it ibid.} D {\bf 76}, 016004 (2007)
  [arXiv:hep-ph/0610385].

\bibitem{Whiteson10}
  C.J.~Flacco, D.~Whiteson, T.M.P.~Tait and S.~Bar-Shalom,
  %``Direct Mass Limits for Chiral Fourth-Generation Quarks in All Mixing
  %Scenarios,''
  Phys.\ Rev.\ Lett.\  {\bf 105}, 111801 (2010)
  [arXiv:1005.1077 [hep-ph]];
  C.J.~Flacco, D.~Whiteson and M.~Kelly,
  arXiv:1101.4976 [hep-ph].

\bibitem{Sjostrand:2006za}
  T.~Sj\"ostrand, S.~Mrenna and P.~Skands,
%  ``PYTHIA 6.4 physics and manual,''
  JHEP {\bf 0605}, 026 (2006)
  [arXiv:hep-ph/0603175].
  %%CITATION = JHEPA,0605,026;%%



\end{thebibliography}
\end{document}